\documentclass[11pt]{article}
\usepackage{iap2000,epsfig}

\def\Journal#1#2#3#4{{#1} {\bf #2}, #3 (#4)}

\def\be{\begin{equation}}
\def\ee{\end{equation}}
\def\bea{\begin{eqnarray}}
\def\eea{\end{eqnarray}}

\def\asp{\em ASP Conf. Ser.}
\def\aj{\em AJ}
\def\apj{\em Apj}
\def\apjs{\em ApjS}
\def\aap{\em A\&A}
\def\aaps{\em A\&AS}
\def\mnras{\em MNRAS}
\newcommand {\h} {$h^{-1} \, Mpc \,$}
\newcommand {\hh} {$h^{-1} \, Mpc$}
\newcommand {\ks} {$km~s^{-1} \;$}

\def\lesssim{\mathrel{\hbox{\rlap{\hbox{\lower4pt\hbox{$\sim$}}}\hbox{$<$}}}}
\def\gtrsim{\mathrel{\hbox{\rlap{\hbox{\lower4pt\hbox{$\sim$}}}\hbox{$>$}}}}
\begin{document}
\title{DYNAMICAL EVOLUTION OF CLUSTERS
\\[16pt]
%\mbox{\includegraphics[]{girardi.eps}}}
\mbox{\epsfysize=5cm \epsffile{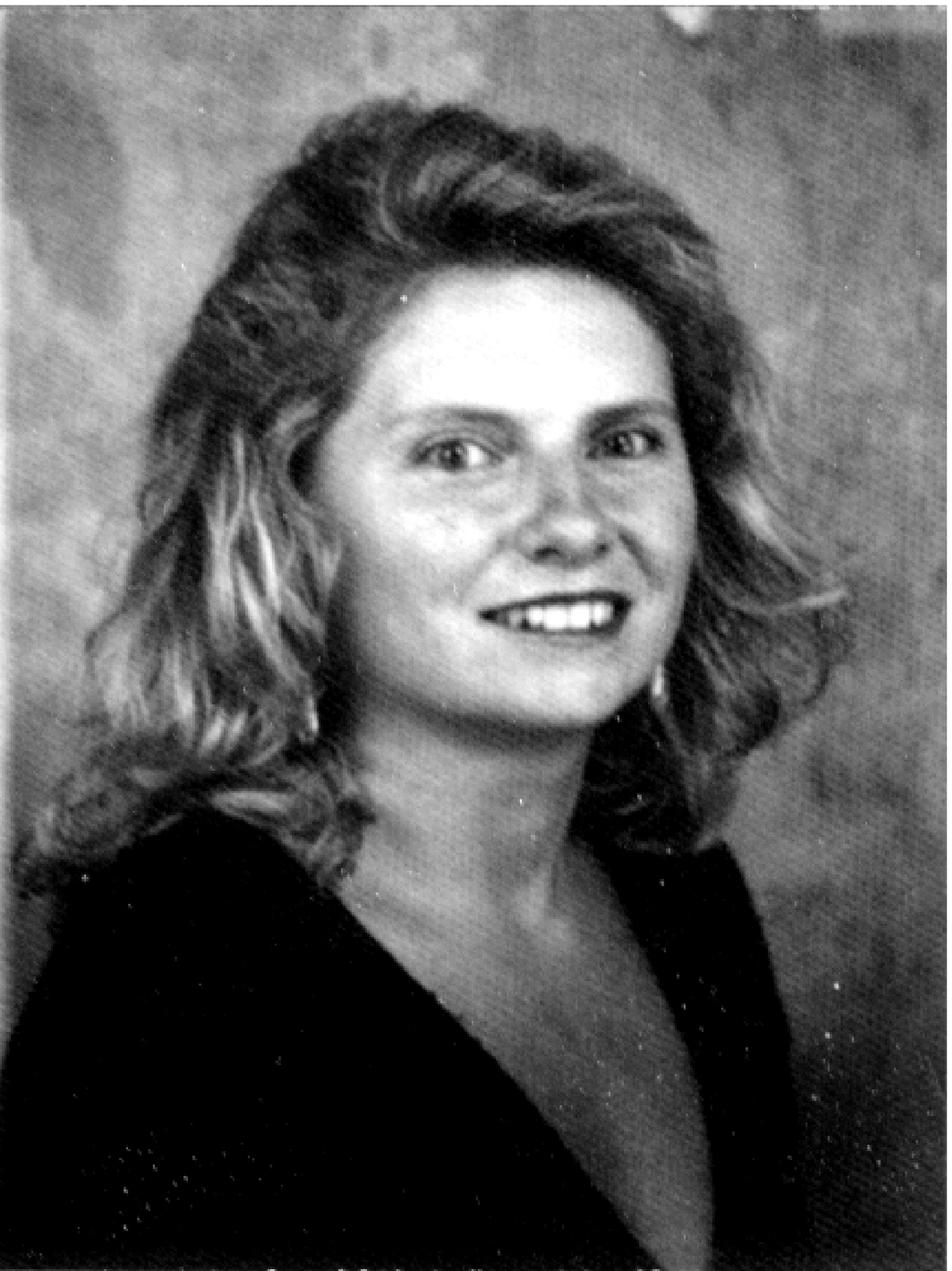}}}

\author{ M. GIRARDI, M. MEZZETTI, and A. DEVETAK }

\address{
Dipartimento di Astronomia, Universit\'a di Trieste (DAUT) \\
via Tiepolo 11, I-34100, Trieste, Italy}

\maketitle\abstracts{
We consider a sample of about 50 distant galaxy clusters at $z>0.15$
($<z>\sim0.3$), each cluster having at least 10 galaxies with
available redshift in the literature.  We select member galaxies,
analyze the velocity dispersion profiles, and evaluate in a
homogeneous way cluster velocity dispersions and virial masses.  We
apply the same procedures already recently applied on a sample of
nearby clusters ($z<0.15$) in order to properly analyze the possible
dynamical evolution of galaxy clusters.  We do not find any
significant difference between nearby and distant clusters.  In
particular, we consider the galaxy spatial distribution, the shape of
the velocity dispersion profile, and the relations between velocity
dispersion and X--ray luminosity and temperature.  Our results imply
little dynamical evolution in the range of redshift spanned by our
cluster sample, and suggest that the typical redshift of cluster
formation is higher than that of the sample we analyze.}

\section{Introduction}
The knowledge of the properties of galaxy clusters plays an important
role in the study of large scale structure formation. In particular,
one can use their number density as a function of their mass ("mass
function") to constrain cosmological parameters. Several recent works
attempt to estimate the value of the matter density parameter
$\Omega_m$ from the evolution of cluster mass function (e.g., 
Carlberg {\it et al.}~\cite{carl97};
Borgani {\it et al.}~\cite{borg99}).

However, the estimate of cluster masses is not an easy task.  Both the
virial theorem applied to positions and velocities of cluster member
galaxies and the dynamical analysis of hot X-ray emitting gas assume
that clusters are systems in dynamical equilibrium, while the mass
estimates derived from gravitational lensing phenomena require a good
knowledge of cluster geometry.

As for the analysis of the internal dynamics of nearby clusters (at redshift
$z\lesssim 0.15$), recent conclusions are based on very large samples
(Fadda {\it et al.}~\cite{fad96}; 
Mazure {\it et al.}~\cite{maz96}; Girardi {\it et al.}~\cite{gir98b}).
In particular, Girardi {\it et al.}~\cite{gir98b} showed that virial masses 
based on member galaxies are
quite consistent with those obtained from X-ray analysis suggesting
that most clusters are not very far from dynamical equilibrium.

As for distant clusters, most results come from the analysis of the 16
clusters at intermediate redshifts, $0.18<z<0.55$ of CNOC (Canadian
Network for Observational Cosmology; Yee {\it et al.}~\cite{yee96}) 
which represents a remarkably homogeneous sample.  In
particular, as found in nearby clusters, Lewis {\it et al.}~\cite{lew99} 
claim
for consistency between masses coming from optical and X--ray data.
However, the difficulties of obtaining many redshifts in distant
clusters have prevented from building larger samples. Rather, several
works, concerning one or a small number of clusters, and using
different techniques of analysis, can be found in the literature.

The availability of a variety of techniques, already applied to nearby
clusters, suggests their application to distant clusters.  We thus
ensure the homogeneity of our results over a large range of
cosmological distances.  A homogeneous analysis is in fact fundamental
for the understanding of the evolution of cluster properties.
We present the preliminary results of our analysis of a sample of
about 50 clusters with $z>0.15$ where we use the same 
techniques already used by Girardi {\it et al.}~\cite{gir98b} (cf. also 
Fadda {\it et al.}~\cite{fad96}) on a
sample of 170 nearby clusters (at $z<0.15$, data from ENACS 
-- ESO Nearby Abell Cluster Survey, Katgert {\it et al.}~\cite{kat98} -- and other
literature).

Unless otherwise stated, we give errors at the 68\% confidence level
(hereafter c.l.).  A Hubble constant of 100 $h$ \ks $Mpc^{-1}$ and a
deceleration parameter of $q_0=0.5$ are used throughout.

\section{The Data Sample}

We analyze a sample of 51 distant galaxy clusters ($z>0.15$,
$<z>=0.3$), each cluster having at least 10 galaxies with available
redshift in the literature, for a total of $\sim 3500$ galaxies.
The sample is a compilation of published data (cf. Girardi \& Mezzetti
in preparation for the complete reference list).

In order to select member galaxies, we apply the same procedure as
Girardi {\it et al.}~\cite{gir98b} (cf. also Fadda et
al.~\cite{fad96}). First, we use the adaptive kernel technique by
Pisani~\cite{pis93} as described by Girardi
et al.~\cite{gir96} to
find the significant peaks in velocity distributions.  Then, we use
the combination of position and velocity information to identify
possible interlopers in the above-detected systems.  We apply the
procedure of the ``shifting gapper'', i.e.  we apply the fixed gap
method to a bin shifting along the distance from the cluster center
(cf. Fadda {\it et al.}~\cite{fad96}).

Finally, we reject galaxies which show strong emission lines. In fact,
there are evidences that emission line galaxies enhance the observed
velocity dispersion, $\sigma_v$, suggesting that these galaxies are
not in dynamical equilibrium within the cluster (e.g., 
Biviano {\it et al.}~\cite{biv97}).

We find that 45 cluster fields show only one peak in their velocity
distribution, and three fields show two separable peaks, for a total
of 51 well separated systems. The other three cluster fields show two
strongly superimposed peaks which suggest that their dynamics is
strongly uncertain. In the following analyses we consider only the 51
well separated systems.
These 51 systems are
those used in the comparison with nearby clusters (160 well defined
systems, cf. Girardi et al.~\cite{gir98b}).

\section{Internal Dynamics}

We estimate the ``robust'' velocity dispersion line--of--sight,
$\sigma_v$, by using the biweight and the gapper estimators when the
galaxy number is larger and smaller than 15, respectively (cf. ROSTAT
routines -- see Beers {\it et al.}~\cite{bee90}), and applying the
relativistic correction and the usual correction for velocity errors
(Danese {\it et al.}~\cite{dan80}).

%\vspace{-10cm}
\begin{figure}
%\rule{5cm}{0.2mm}\hfill\rule{5cm}{0.2mm}
%\vskip 2.5cm
%\rule{5cm}{0.2mm}\hfill\rule{5cm}{0.2mm}
\hspace{-2cm}
\psfig{figure=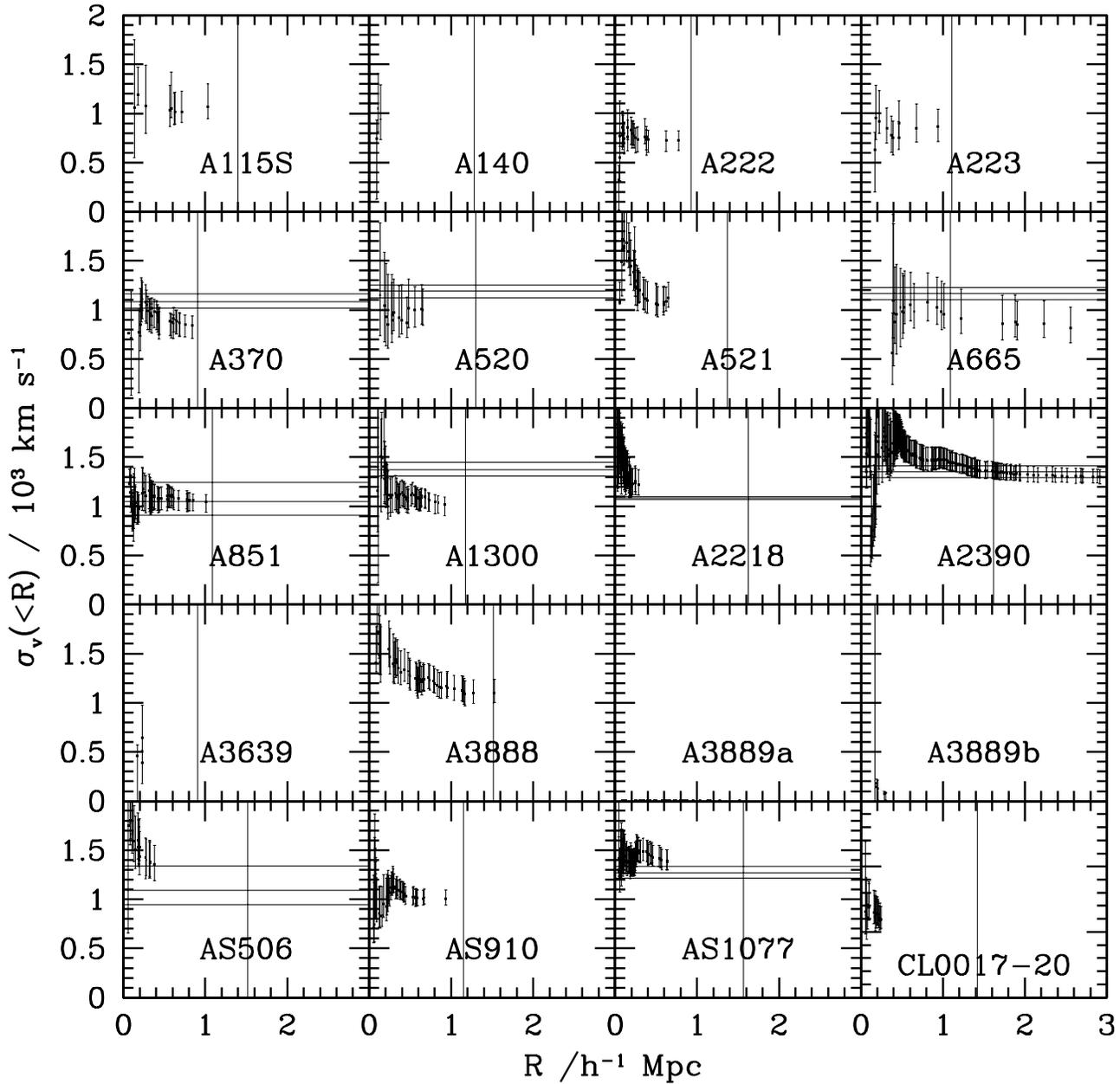,height=18cm}
\caption{Integrated line--of--sight velocity dispersion profiles
$\sigma_v(<R)$, where the dispersion at a given (projected) radius
from the cluster center is estimated by considering all galaxies
within that radius. The bootstrap error bands at the $68\%$ c.l. are
shown.  
The horizontal lines represent X--ray temperature with the respective
errors transformed in $\sigma_v$ imposing $\beta_{spec}=1$ (cf. \S~4).  
The vertical faint line indicates the virialized region
within $R_{vir}$.
\label{fig:f1}}
\end{figure}

Following Fadda {\it et al.}~\cite{fad96} (cf. also Girardi {\it et
al.}~\cite{gir96}) we analyze the ``integral'' velocity dispersion
profile (hereafter VDP), where the dispersion at a given (projected)
radius is evaluated by using all the galaxies within that radius,
i.e. $\sigma_v(<R)$. The VDP allows to check the robustness of
$\sigma_v$ estimate. In particular, although the presence of velocity
anisotropies can strongly influence the value of $\sigma_v$ computed
for the central cluster region, it does not affect the value of the
$\sigma_v$ computed for the whole cluster (e.g.,
Merritt~\cite{mer88}). The VDPs of nearby clusters show strongly
increasing or decreasing behaviors in the central cluster regions, but
they are flattening out in the external regions (beyond $\sim $1 \hh,
cf. also den Hartog \& Katgert~\cite{den96}) suggesting that in such
regions they are no longer affected by velocity anisotropies.  Thus,
while the $\sigma_v$-values computed for the central cluster region
could be a very poor estimate of the depth of cluster potential wells,
one can reasonably adopt the $\sigma_v$ value computed by taking all
the galaxies within the radius at which the VDP becomes roughly
constant.

As for the distant clusters we analyze, when the data are good enough,
the VDPs show a behavior similar to that of nearby clusters
(cf. VDPs for 24 out of 51 clusters in Figure~1).  
Unfortunately, distant clusters suffer for the poor
sampling, and also for the small spatial extension of the sampled
cluster region.  Indeed, the strongly decreasing VDP in the external
sampled regions of some clusters (cf. AS506 in Figure~1) 
suggests that the correct estimates of
velocity dispersions could be smaller than those, $\sigma_v$, we can
estimate with present data; therefore, in these cases, $\sigma_v$
should be better interpreted as an upper limit. In other cases,
when the member galaxies are too few, the analysis of VDPs does not
allow any conclusion.

After fixing the cosmological background, the theory of a spherical model
for nonlinear collapse allows to recover the value of the radius of
virialization, $R_{vir}$, within which the cluster can considered not far
from a status of dynamical equilibrium.  
For nearby clusters Girardi {\it et al.}~\cite{gir98b} give a first roughly estimate of
$R_{vir}\sim 0.002 \cdot \sigma_v$ ($km^{-1}s\ $ \hh). A following
re--estimate of Girardi {\it et al.}~\cite{gir98a} suggests rather a scaling factor of
0.0017. Since we find that distant clusters have a galaxy distribution
similar to that of nearby ones (see in the following), we adopt here the
same scaling relation with $\sigma_v$: i.e.
$R_{vir}\sim 0.0017 \cdot \sigma_v/(1+z)^{3/2} \,\, (km^{-1} s\  h^{-1}\,Mpc)$ 
introducing only the scaling with redshift (cf. also
Carlberg {\it et al.}~\cite{carl97c} for a similar relation).

We analyze galaxy distribution in a similar way to that used by 
Girardi {\it et al.}~\cite{gir98b},
i.e. by fitting the galaxy surface density of each cluster to a King
distribution with a variable exponent (hereafter referred to as a
``King-like'' profile, cf. Girardi {\it et al.}~\cite{gir95}):
$\Sigma(R)=\Sigma_0/(1+(R/R_c)^2)^{\alpha}$, where $R_c$ is the core
radius and $\alpha$ is the parameter which describes the galaxy
distribution in external regions.  This surface density profile
corresponds to a galaxy volume-density $\rho(r) \propto
r^{-(2\alpha+1)}$ for $r>> R_C$.  We perform the fit through the
Maximum Likelihood technique, allowing $R_C$ and $\alpha$ to vary from
0.01 to 1 \h and from 0.5 to 1.5, respectively. We perform the fit
within the circular cluster region, of radius $R_{max,c}$, all
contained within the sampled cluster region. We consider only the 30
clusters with at least ten member galaxies within $R_{max,c}$ and we
verify our results on a subsample of 13 clusters with
$R_{max,c}/R_{vir}>0.5$.

The median value of $\alpha$, with the respective errors at the $90\%$
c.l., is $=0.63_{-0.08}^{+0.08}$. This value agrees with
$\alpha=0.70_{-0.03}^{+0.08}$ found for nearby clusters, and
corresponds to a $\beta_{fit,gal}\sim0.8$, i.e. to a volume
galaxy--density $\rho \propto r^{-2.4}$.  After fixing $\alpha=0.7$,
we again fit the galaxy distribution of each cluster, obtaining a
median value of $R_c=0.045_{-0.015}^{+0.005}$ \h. Thus, in our cluster
sample, the typical value of $R_c$ (and $R_{vir}/R_c \sim 20$) is
again in agreement with that found in nearby clusters where
$R_c=0.05\pm0.01$ \hh.  Hereafter, we assume the above King--modified
distribution, with the same parameters of nearby clusters, i.e.
$\alpha=0.7$ and $R_c=0.05$ \hh, for all clusters of our sample.

The standard methods used to estimate the cluster mass from member galaxies
require that galaxies are in equilibrium within the cluster potential.
The cluster mass is then recovered from the knowledge of positions and
velocities of the same population of galaxies which are taken as
tracers of the cluster potential.  

Assuming that clusters are spherical, non rotating systems, and that
the internal mass distribution follows galaxy distribution, cluster
masses can be computed throughout the virial theorem (e.g., 
Limber \& Mathews~\cite{lim60}; The \& White~\cite{the86}) as:
 
\begin{equation}
M=M_V-C=\frac{3\pi}{2} \cdot \frac{\sigma_v^2 R_{PV}}{G}-C,
\label{eq:massa}
\end{equation}

\noindent where the projected virial radius,
$R_{PV}=N(N-1)/(\Sigma_{i> j} R_{ij}^{-1})$, describes the galaxy
distribution and is computed from projected mutual galaxy distances,
$R_{ij}$; $C$ is the surface term correction to the standard virial
mass $M_V$ and it is due to the fact that the system is not entirely
enclosed in the observational sample (cf. also 
Carlberg {\it et al.}~\cite{carl96};
Girardi {\it et al.}~\cite{gir98b}).

\begin{figure}
%\rule{5cm}{0.2mm}\hfill\rule{5cm}{0.2mm}
%\vskip 2.5cm
%\rule{5cm}{0.2mm}\hfill\rule{5cm}{0.2mm}
\hspace{2cm}
\psfig{figure=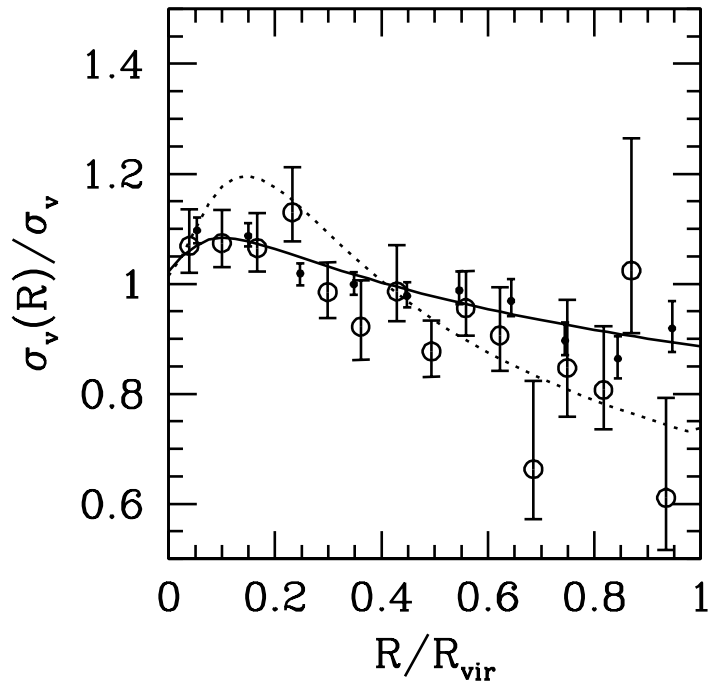,height=20cm}
\vspace{-10cm}
\caption{
The (normalized) line--of--sight velocity
dispersion, $\sigma_v(R)$, as a function of the (normalized) projected
distance from the cluster center.  The points represent data combined
from all clusters and binned in equispatial intervals. We give the
robust estimates of velocity dispersion and the respective bootstrap
errors.  We give the results for distant clusters (open circles) and
for nearby clusters taken from Girardi {\it et al.}$^5$ (filled circles).
The solid and dotted line represent the models for isotropic and
moderate radial orbits of galaxies, respectively.
\label{fig:f2}}
\end{figure}

Following Girardi {\it et al.}~\cite{gir98b} 
we want to estimate cluster masses contained within the
radius of virialization, $R_{vir}$.  In fact, clusters cannot be
assumed in dynamical equilibrium outside $R_{vir}$ and considering
small cluster region leads to unreliable measure of the potential
($\sigma_v$ could be strongly affected by velocity anisotropies) and
of the surface term correction 
(Koranyi \& Geller~\cite{kor00}).

Unfortunately, only few distant clusters are sampled out to $R_{vir}$.
As for $\sigma_v$, the above analysis of the VDP give indications
about its reliability, i.e. VDPs which are flat in the external
cluster regions will give reliable estimates of $\sigma_v$.  As for
$R_{PV}$, which describes the galaxy spatial distribution, it can be
recovered in an alternative theoretical way from the knowledge of the
parameters of the King--like distribution (Girardi {\it et al.}~\cite{gir95}; see
also Girardi {\it et al.}~\cite{gir98b} 
for a simple analytical approximation in the case of
$\alpha=0.7$ and $R_c/R_{vir}=0.05$).  One can compute $R_{PV}$ at
each cluster radius and, in particular, we compute $R_{PV}$ at
$R_{vir}$, which is needed in the computation of the mass within
$R_{vir}$.

The computation of the $C$ correction at the boundary radius, here $R_{vir}$,
is given in eq.~14 of Girardi {\it et al.}~\cite{gir98b} and 
requires the knowledge of the velocity anisotropy of galaxy
orbits.

Having assumed that in clusters the mass distribution follows the
galaxy distribution, one can use the Jeans equation to estimate
velocity anisotropies from the data, i.e. from the (differential)
profile of the line--of--sight velocity dispersion, $\sigma_v(R)$.  We
compute the observational $\sigma_v(R)$ by combining together the
galaxies of all clusters, i.e.  by normalizing distances to $R_{vir}$
and velocities, relative to the mean cluster velocity, to the observed
global velocity dispersion $\sigma_v$. For nearby clusters the
observational profile is well described by a theoretical profile
obtained by the Jeans equation, assuming that velocities are
isotropic, i.e. that the tangential and radial components of velocity
dispersion are equal (i.e., the velocity anisotropy parameter ${\cal
A}=1-\sigma_{\theta}^2(r)/\sigma_r^2(r)=0$).  For distant clusters
this model is less satisfactory (Figure~2), but cannot be rejected
being acceptable at the $\sim15\%$ c.l. (according to the $\chi^2$
probability).

In order to give $C$--corrections more appropriate to each individual
cluster Girardi {\it et al.}~\cite{gir98b} used a profile indicator, $I_p$, which is the ratio
between $\sigma_v(<0.2\times R_{vir})$, the line--of--sight velocity
dispersion computed by considering the galaxies within the central
cluster region of radius $R=0.2\times R_{vir}$, and the global
$\sigma_v$. For 33 clusters we can compute the profile indicator and
the relative correction; for 18 clusters we cannot define the kind of
profile and we assume isotropic orbits ($20\%$ of correction).

\section{Comparison with X-ray and Lensing Results}

We collect X-ray luminosities, in general bolometric ones,
$L_{bol,X}$, and temperature, $T_X$, for 38 and 22 clusters,
respectively (most of the data coming from the compilation of 
Wu {\it et al.}~\cite{wu99}; 
cf. Girardi \& Mezzetti in preparation for the complete reference
list).

\begin{figure}
%\rule{5cm}{0.2mm}\hfill\rule{5cm}{0.2mm}
%\vskip 2.5cm
%\rule{5cm}{0.2mm}\hfill\rule{5cm}{0.2mm}
\hspace{3cm}
\psfig{figure=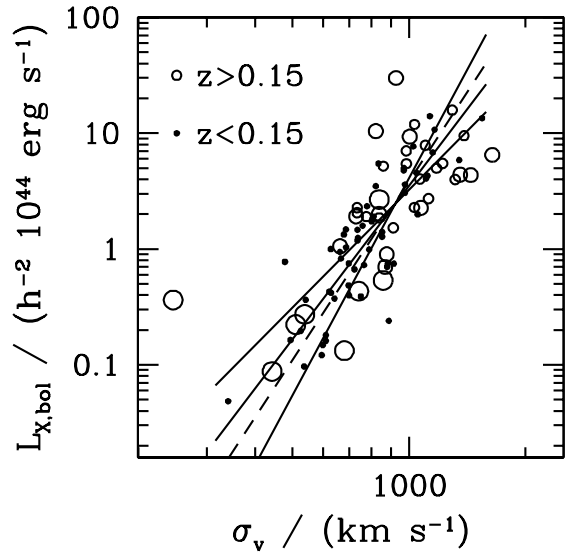,height=16cm}
\vspace{-8.5cm}
\caption{ $L_{X,bol}$--$\sigma_v$ relation for distant (open circles)
and nearby clusters (filled circles).  For the distant clusters, the
circle size decreases with the number of galaxies used to estimate
$\sigma_v$: the smallest, the intermediate, and the largest circles
indicate $N_m\ge30$, $10\le N_m<30$, and $N_m<10$, respectively.  For
the nearby clusters we show results as reported by Borgani {\it et
al.}$^{2}$, all having $\sigma_v$ estimated at least with 30
galaxy redshifts (Girardi {\it et al.}$^{5}$) and also
belonging to the X--ray Brightest Abell--like Cluster survey (Ebeling
{\it et al.}$^{24}$).  The three solid lines are direct,
inverse, and bisecting linear regression for the distant clusters
(obtained rejecting the point on the left).  The dashed line is the
bisecting linear regression for the nearby clusters as computed by
Borgani {\it et al.}$^{2}$.
\label{fig:f3}}
\end{figure}

Figure~3
shows the $L_{X,bol}$--$\sigma_v$ relation compared to that
found by Borgani {\it et al.}~\cite{borg99} for nearby clusters.  
Excluding the leftmost point
(J2175.15TR), the resulting bisecting linear regression is

\begin{equation}
log(L_{bol,X}/10^{44}erg\,s^{-1})=4.4^{+1.8}_{-1.0}\,log(\sigma_v/km \,s^{-1})-12.6^{+3.0}_{-5.4}\,,
\label{eq:lum}
\end{equation}

\noindent where errors come from the difference with respect to the
direct and the inverse linear regression (Isobe {\it et al.}~\cite{iso90}, OLS
methods).  Our $L_{bol,X}$--$\sigma_v$ relation is consistent with
that of nearby clusters (e.g., White {\it et al.}~\cite{whi97}; Borgani
{\it et al.}~\cite{borg99};
Wu {\it et al.}~\cite{wu99}).  As for the point excluded, note that our
analysis of J2175.15TR is based only on 19 galaxies, and the estimate
of $\sigma_v$ is recovered from only eight member galaxies (with an
error larger than $100\%$).

\begin{figure}
%\rule{5cm}{0.2mm}\hfill\rule{5cm}{0.2mm}
%\vskip 2.5cm
%\rule{5cm}{0.2mm}\hfill\rule{5cm}{0.2mm}
\hspace{3cm}
\psfig{figure=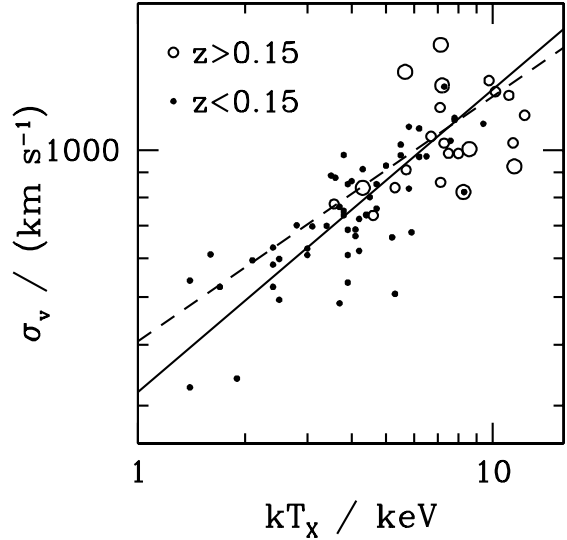,height=16cm}
\vspace{-8.5cm}
\caption{
$\sigma_v$--$T_X$ relation for distant (open
circles) and nearby clusters (filled circles).  For the distant
clusters, the circle size decreases with the number of galaxies used
to estimate $\sigma_v$: the smaller and the larger circles indicate
$N_m\ge 30$, and $10\le N_m<30$, respectively.  For the nearby clusters
we show results as reported by Girardi {\it et al.}$^5$, all having
$\sigma_v$ estimated at least with 30 galaxy redshifts, and
$T_X$ taken from David {\it et al.}$^{27}$ and from 
White {\it et al.}$^{26}$.  
The solid
line is the bisecting linear regression for the nearby clusters as
computed by Girardi {\it et al.}$^5$.  
The dashed line represents the model with
the equipartition of energy per unit mass between gas and galaxy
components ($\beta_{spec}=1$).
\label{fig:f4}}
\end{figure}

Figure~4 shows the $\sigma_v$--$T_X$ relation compared to that of
nearby clusters, as reported by Girardi et al.~\cite{gir98b}.  As for
distant clusters, the data have a too small dynamical range to attempt
a linear fit: the visual inspection of Figure~4 suggests no difference
with nearby clusters in agreement with the result of by Mushotzky \&
Scharf~\cite{mus97} and Wu {\it et al.}~\cite{wu99}.  We obtain
$\beta_{spec}= \sigma_v^2/(kT/\mu m_p)= 0.88^{+0.14}_{-0.17}$, where
$\mu=0.58$ is the mean molecular weight and $m_p$ the proton mass
(median value with errors at the $90\%$ c.l.).
This value of $\beta_{spec}$ is in good agreement with the
value of $\beta_{spec}=0.88\pm0.04$ for nearby clusters (cf.  Girardi
{\it et al.}~\cite{gir98b}).  Moreover, we find no correlation between
$\beta_{spec}$ and redshift (cf. also Wu {\it et al.}~\cite{wu99}).

Under the assumption that the hot diffuse gas is in hydrostatic and
isothermal equilibrium with the underlying gravitational potentials of
clusters, one can obtain the X--ray cluster masses provided that the
gas temperature and radial profile of gas distribution are known.
The availability of $T_X$ allow us to compute
the mass within $R_{vir}$ for 22 clusters according to
$M_{X}=(3\beta_{fit,gas}kT\cdot R_{vir})/(G \mu m_p) \cdot
(R_{vir}/R_x)^2/[1+(R_{vir}/R_x)^2]$, where we adopt the gas
distribution given by the $\beta$--model with typical parameters
(slope $\beta_{fit,gas} =2/3$ and core radius $R_x=0.125$ \h,
e.g., Jones \& Forman~\cite{jon92}).  We find mass values consistent with our
optical virial estimates, i.e.  $M/M_{X}=1.02\,(0.86$--$1.32)$ for the
median value and the range at the $90\%$ c.l..

As for gravitational lensing masses, we resort to estimates found in
the literature. We collect projected estimates from weak gravitational
lensing analysis, $M_L$, for 18 clusters (cf. Girardi \& Mezzetti in
preparation for the complete reference list).  In order to compare our
optical virial masses to $M_L$, we project and rescale our masses $M$
within the corresponding radius using the fitted galaxy spatial
distribution.  We obtain $M_{opt,L}/M_L=1.30(0.63$--$2.13)$ (median
value and range at the $90\%$ c.l.).  Moreover, we do not find any
correlation between $M/M_X$ or $M_{opt,L}/M_L$ and redshift.

Our finding are in agreement with other recent studies which find, on
average, no evidence of discrepancy between different mass estimates
as computed within large radii, thus suggesting that distant clusters
are nor far from global dynamical equilibrium (e.g., Allen~\cite{all98}; 
Lewis {\it et al.}~\cite{lew99}).  Note that we avoid to consider mass
determination in very central cluster regions since our analysis of
cluster members give poor constrains on mass distribution on these
scales.  Indeed, the assumption of dynamical equilibrium seem to break
down in the very central regions as suggested by comparisons with
strong lensing mass estimates (e.g., Allen~\cite{all98};
Lewis {\it et al.}~\cite{lew99}).

\section{Summary and Conclusions}

In order to properly analyze the possible dynamical evolution of
galaxy clusters we apply the same procedures already applied on a
sample of nearby clusters (170 clusters at $z<0.15$ from ENACS and
other literature, Girardi {\it et al.}~\cite{gir98b}, cf. also Fadda {\it et al.}~\cite{fad96}) to
a corresponding sample of distant clusters.

We consider a sample of 51 distant galaxy clusters at $z>0.15$
($<z>\sim0.3$), each cluster having at least 10 galaxies with
available redshift in the literature.  A part from three cluster field
showing two overlapping peaks in their velocity distribution and so
large uncertainties in their dynamics, 45 fields show only one peak in
their velocity distribution and three fields show two separable peaks
for a total of 51 well defined cluster systems.  
These 51 systems are
those used in the comparison with nearby clusters (160 well defined
systems).

We select member galaxies, analyze the velocity dispersion profiles,
and evaluate in a homogeneous way cluster velocity dispersions and
virial masses.

As a main general result, we do not find any significant evidence for
dynamical evolution of galaxy clusters. More in detail, our results
can be summarized as follows.

\begin{itemize}

\item The galaxy spatial distribution is similar to that of nearby
clusters, i.e. the fit to a King--like profile gives a two-dimensional
slope of $\alpha=0.7$ and a very small core radius of $R_c=0.05$
\hh. 

\item When data are good enough, the integrated velocity dispersion
profiles of distant clusters show a behavior similar to those of
nearby clusters, i.e.  they show strongly increasing or decreasing
behaviors in the central cluster regions, but are flattening out in
the external regions suggesting that in such regions, they are no
longer affected by velocity anisotropies.

\item The average velocity dispersion profile can be explained by a
model with isotropic orbits, which well describe also nearby clusters.
Possible evidences for more radial orbits are not statistically
significant.

\item There is no evidence of evolution in both
$L_{bol,X}$--$\sigma_v$ and $\sigma_v$--$T_X$ relations, thus in
agreement with previous results (Mushotzky \& 
Scharf~\cite{mus97}; Borgani {\it et al.}~\cite{borg99}).

\end{itemize}

Moreover, on average, within the large scatter of present data, we
find no significant evidence of discrepancies between our virial mass
estimates and those from X--ray and gravitational lensing data, thus
suggesting that distant clusters are not far from global dynamical
equilibrium (cf. also Allen~\cite{all98}; Lewis {\it et al.}~\cite{lew99}).

We conclude that the typical redshift of cluster formation is higher
than that of our sample in agreement with previous suggestions (e.g.,
Schindler~\cite{sch99}; Mushotzky~\cite{mus00}).  In particular,
we agree with preliminary results by Adami {\it et al.}~\cite{ada99}, who applied
the same techniques used for the nearby ENACS clusters on 15 distant
clusters, ($<z>\sim0.4$) from the Palomar Distant Cluster Survey
(Postman {\it et al.}~\cite{pos96}).

Although some clusters at very high redshift, e.g. $z>0.8$, are
already known (e.g., Gioia {\it et al.}~\cite{gio99}; Rosati 
{\it et al.}~\cite{ros99}), the
construction of a large cluster sample useful for studying internal
dynamics will require a strong observational effort. We stress how
both the poor number of galaxies and the small spatial extension of
some clusters can affect the robustness of their resulting properties.


\section*{References}
\begin{thebibliography}{99}

\bibitem{carl97} R.G. Carlberg {\it et al.}, \Journal{\apj}{476}{L7}{1997}

\bibitem{borg99} S. Borgani {\it et al.},
\Journal{\apj}{527}{561}{1999}

\bibitem{fad96} D. Fadda {\it et al.}, 
\Journal{\apj}{473}{670}{1996}

\bibitem{maz96} A. Mazure {\it et al.}, \Journal{\aap}{310}{31}{1996}

\bibitem{gir98b} M. Girardi
{\it et al.}, \Journal{\apj}{505}{74}{1998}

\bibitem{yee96} H.K.C. Yee
{\it et al.},
\Journal{\apjs}{102}{269}{1996}

\bibitem{lew99} A.D. Lewis
{\it et al.},
\Journal{\apj}{517}{587}{1999}

\bibitem{kat98} P. Katgert
{\it et al.},
\Journal{\aaps}{129}{399}{1998}

\bibitem{pis93} A. Pisani,
\Journal{\mnras}{265}{706}{1993}

\bibitem{gir96} M. Girardi
{\it et al.},
\Journal{\apj}{457}{61}{1996}

\bibitem{biv97} A. Biviano
{\it et al.},
\Journal{\aap}{321}{84}{1997}

\bibitem{bee90} T.C. Beers
{\it et al.}, \Journal{\aj}{100}{32}{1990}

\bibitem{dan80} L. Danese
{\it et al.},
\Journal{\aap}{82}{322}{1980}

\bibitem{mer88} D. Merritt,
\Journal{\asp}{5}{175}{1988}

\bibitem{den96} R. den Hartog and P. Katgert, 
\Journal{\mnras}{279}{349}{1996}

\bibitem{gir98a} M. Girardi
{\it et al.},
\Journal{\apj}{506}{45}{1998}

\bibitem{carl97c} R.G. Carlberg
{\it et al.}, \Journal{\apj}{478}{462}{1997}

\bibitem{gir95} M. Girardi
{\it et al.}, \Journal{\apj}{438}{527}{1995}

\bibitem{lim60} D.N. 
Limber and W.G. Mathews
{\it et al.}, \Journal{\apj}{132}{286}{1960}

\bibitem{the86} L.S. The and  S.D.M. White,
\Journal{\aj}{92}{1248}{1960}

\bibitem{carl96} R.G. Carlberg
{\it et al.},
\Journal{\apj}{462}{32}{1996}

\bibitem{kor00} D.M. Koranyi and M.J. Geller,
\Journal{\aj}{119}{44}{2000}

\bibitem{wu99} X.P. Wu
{\it et al.},
\Journal{\apj}{524}{22}{1999}

\bibitem{ebe96} H. Ebeling,
\Journal{\mnras}{281}{799}{1996}

\bibitem{iso90} T. Isobe
{\it et al.},
\Journal{\apj}{364}{104}{1990}

\bibitem{whi97} D.A. White
{\it et al.},
\Journal{\mnras}{292}{419}{1997}

\bibitem{dav93} L. P. David
{\it et al.},
\Journal{\apj}{412}{479}{1993}

\bibitem{mus97} R.F. Mushotzky and C.A. Scharf,
\Journal{\apj}{482}{L13}{1997}

\bibitem{jon92} C. Jones and W. Forman, 
in {\em Clusters and Superclusters
of Galaxies}, ed. A. C. Fabian (Dordrecht: Kluwer), p. 49 (1992)

\bibitem{all98} S.W. Allen, 
\Journal{\mnras}{296}{392}{1998}

\bibitem{sch99} S. Schindler, 
\Journal{\aap}{349}{435}{1999}

\bibitem{mus00} R.F. Mushotzky, 
\Journal{\asp}{193}{323}{2000}

\bibitem{ada99} C. Adami
{\it et al.},
in {\em proc. of
the 1999 IGRAP Conference}, 
preprint astro-ph/9907366 (1999)

\bibitem{pos96} M. Postman
{\it et al.},
\Journal{\aj}{111}{615}{1996}

\bibitem{gio99} I.M. Gioia
{\it et al.},
\Journal{\aj}{117}{2608}{1999}

\bibitem{ros99} P. Rosati
{\it et al.},
\Journal{\aj}{118}{76}{1999}

\end{thebibliography}
\end{document}